# On the robustness of the *h*-index

Jerome K. Vanclay

School of Environmental Science and Management

Southern Cross University, Lismore NSW 2480, Australia

Tel +61 2 6620 3147, Fax +61 2 6621 2669, JVanclay@scu.edu.au

**Abstract**

The *h*-index (Hirsch, 2005) is robust, remaining relatively unaffected by errors in the long tails of the citations-rank distribution, such as typographic errors that short-change frequently-cited papers and create bogus additional records. This robustness, and the ease with which h-indices can be verified, support the use of a Hirsch-type index over alternatives such as the journal impact factor. These merits of the *h*-index apply to both individuals and to journals.

**Introduction**

Despite well-recognised flaws (e.g., Jennings 1998; Seglen 1997), the ISI journal impact factor (JIF, the mean number of citations per paper) continues to have a major influence on scientific endeavour (Bordons et al 2002; Monastersky 2005). Hirsch (2006) proposed an alternative *h*-index that has been shown to be effective (Bornmann & Daniel 2005; Oppenheim 2006) and consistent with other metrics (Cronin and Meho 2006). Although initially proposed for individual scientists, others have suggested extensions of the *h*-index to teams and journals (e.g. Braun et al

2005). However, some of the statistical properties of these metrics have not received sufficient attention. Hirsch's *h*-index avoids several problems with the JIF, including censorship (in the statistical sense of truncating data contributing to the numerator or denominator; Butler and Visser 2006), errors (Lange 2002; Gehanno 2005), manipulation (Agrawal 2005; Karandikar and Sunder 2003; Mannino 2005; Monastersky 2005) and with long-tailed distributions (Redner 1998).

League tables usually show impact factors in neat columns with counts of total citations, total publications, and inferred impacts. Sadly, these data are not as precise as they may appear (Garfield 2005, Bensman *in press*). The total number of citations may be affected by error, manipulation, and by the selection of journals and articles that contribute to the count. The total number of publications may also be influenced by censorship (Are editorials included in the published output of a journal? Is 'grey literature' included in the count of an individual's output?). Thus both the number of citations and the number of publications are likely to be approximate and often biased, with the result that the inferred impact factor may include considerable error. These problems of censorship and manipulation are likely to be greatest in both tails of the distribution. For instance, some 'highly-cited' articles may be mentioned in the media or other influential avenues not seen by ISI, while conversely, an arbitrary decision to include (or exclude) contributions in the grey literature may inflate the tally of an individual's total output. Hirsch's *h*-index avoids many of these issues by ignoring the long-tails of the distribution, and focussing on the 'middle part' of the Zipf plot of number of citations versus ranked paper number (Hirsch 2006; Fig. 1).

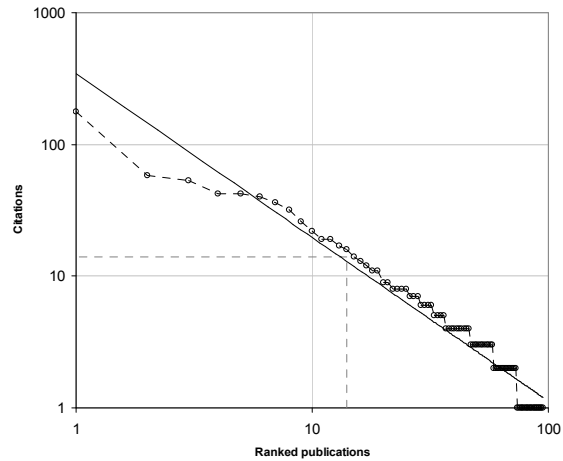

**Figure 1**. Citations accruing to the author's publications, including self-citations and 'grey' publications (conference proceedings, etc). The solid line is a power-curve $Y=aX^b$ and the dashed line indicates *h*-index 14.

Hirsch's *h*-index has two further advantages: it is an integer, so avoids the false impression of precision conveyed by the three decimal points in the ISI impact factor, and is much easier to verify than most alternatives. If disputed, it may be difficult to reliably verify the total number of citations or an index based on the mean number of citations per publication (e.g., the JIF). However, a dispute surrounding a *h*-index is easy to verify. Most of the publications of a journal or individual receive more or many fewer than the *n* citations contributing to a *h*-index of *n*, so verifying the index involves checking the citations accruing to just a few publications ranked higher than *n* (e.g., with *n-1* citations). Such checks simply need to establish whether typographic errors or other factors may have concealed one or two citations associated with these 'threshold' publications, allowing the index to rise to *n+1* after these anomalies are redressed. The great majority of errors (and distortions) in citation databases lie in the long tails, and tend not affect the *h*-index greatly. It is a relatively simple matter to check the citations accruing to one or two publications, in contrast to the challenge of verifying the total number of citations and publications.

**Approach and Methods**

The robustness of the *h*-index is illustrated with my own publication record. Table 1 illustrates the raw data obtained from two service providers (see Bakkalbasi et al 2006 for a comparison of these and other service providers): from Google Scholar (GS) by searching for 'author:j-vanclay' (http://scholar.google.com/scholar?q=author%3Aj-vanclay), and from ISI's Web of Science (WoS) by searching for "VANCLAY J*". A naive interpretation of these raw data (including self-citations) suggests a *h*-index of 11 and 12 respectively, or 13 if based on the larger of these alternatives (Table 1). Both these databases contain some obvious errors (for instance, 3 entries without author-tags in GS, and typographic errors in WoS that generated erroneous duplicates not shown in Table 1). Correcting these obvious errors indicated *h*-indices of 13, 12 and 14 respectively (Table 2).

**Table 1**. Raw citation data retrieved (on 15 May 2006) from Google Scholar (GS) and ISI Web of Science (WoS), including self-citations, truncated at rank 20. Emboldened row indicates the *h*-index for the column based on Max(GS,WoS).

| Rank | GS | WoS | Max | Publication | Date | Vol | Page |
|---|---|---|---|---|---|---|---|
| 1 | 172 | 96 | 172 | Book: Modelling Forest Growth and Yield | 1994 | | |
| 2 | 57 | 50 | 57 | Forest Science | 1995 | 41 | 7 |
| 3 | 53 | 53 | 53 | Ecological Modelling | 1997 | 98 | 1 |
| 4 | 35 | 41 | 41 | Forest Ecology and Management | 1995 | 71 | 267 |
| 5 | 40 | | 40 | Report: A Sustainable Forest Future | 1999 | | |
| 6 | | 40 | 40 | Forest Ecology and Management | 1991 | 42 | 143 |
| 7 | 29 | 36 | 36 | Forest Ecology and Management | 1989 | 27 | 245 |
| 8 | 30 | 32 | 32 | Forest Ecology and Management | 1995 | 71 | 251 |
| 9 | 26 | 10 | 26 | Forest Ecology and Management | 2003 | 172 | 229 |
| 10 | 15 | 19 | 19 | Journal of Tropical Forest Science | 1991 | 4 | 59 |
| 11 | 9 | 17 | 17 | Forest Ecology and Management | 1992 | 54 | 257 |
| 12 | 15 | 16 | 16 | Forest Science | 1991 | 37 | 1656 |
| **13** | **13** | **12** | **13** | **Ambio** | **1993** | **22** | **225** |
| 14 | | 13 | 13 | Forest Ecology and Management | 2001 | 150 | 27 |
| 15 | 11 | 6 | 11 | Forest Ecology and Management | 1994 | 69 | 299 |
| 16 | 11 | 8 | 11 | Canadian Journal of Forest Research | 1992 | 22 | 1235 |
| 17 | 10 | 2 | 10 | Agroforestry Forum | 1998 | 9 | 47 |
| 18 | 10 | 9 | 10 | Forest Ecology and Management | 1997 | 94 | 149 |
| 19 | 6 | 9 | 9 | Photogramm. Eng. and Remote Sensing | 1990 | 56 | 1383 |
| 20 | | 7 | 7 | Forest Ecology and Management | 2001 | 150 | 79 |

Table 2 includes corrections for all 20 entries, but most of these corrections have no bearing on the resulting $h$-index, and it is normally necessary to effect corrections only to entries ranked higher (i.e., with fewer citations) than the preliminary $h$-index. Table 2 assumes that the larger of the two citation counts is a good approximation of the total, but this may not always be so, and it is prudent to examine the union of the two sets of citations. This need be done only for a few cases. Most of the entries in Table 2 already exceed the estimated $h$-index, and a further increase in the citation count will have no bearing on the estimate. And many publications with low citation counts (Figure 1) are unlikely to reach the $h$-index. Thus it is prudent to check citations only for those publications with rank larger than interim $h$-index, and for which the sum of the two citation counts is less than the rank of the publication.

**Table 2**. Citations accruing to top 20 publications after correcting obvious errors. Corrections shown in bold. Italics indicate one publication that increased substantially in rank.

| Rank | GS Raw | GS Correct | WoS Raw | WoS Correct | Max | Publication | Date | Vol | Page |
|---|---|---|---|---|---|---|---|---|---|
| 1 | 172 | **177** | 96 | **142** | 177 | Book: Modelling ... | 1994 | | |
| 2 | 57 | **58** | 50 | **52** | 58 | Forest Science | 1995 | 41 | 7 |
| 3 | 53 | 53 | 53 | 53 | 53 | Ecological Modelling | 1997 | 98 | 1 |
| 4 | 35 | **41** | 41 | **42** | 42 | For. Ecol. Manage. | 1995 | 71 | 267 |
| 5 | 40 | **42** | | | 42 | Report: A Sustainable ... | 1999 | | |
| 6 | | **27** | 40 | 40 | 40 | For. Ecol. Manage. | 1991 | 42 | 143 |
| 7 | 29 | **30** | 36 | 36 | 36 | For. Ecol. Manage. | 1989 | 27 | 245 |
| 8 | 30 | 30 | 32 | 32 | 32 | For. Ecol. Manage. | 1995 | 71 | 251 |
| 9 | 26 | 26 | 10 | 10 | 26 | For. Ecol. Manage. | 2003 | 172 | 229 |
| 10 | 15 | 15 | 19 | 19 | 19 | J. Trop. For. Sci. | 1991 | 4 | 59 |
| *11* | | *19* | *13* | *13* | *19* | *For. Ecol. Manage.* | *2001* | *150* | *27* |
| 12 | 9 | 9 | 17 | 17 | 17 | For. Ecol. Manage. | 1992 | 54 | 257 |
| 13 | 15 | 15 | 16 | 16 | 16 | Forest Science | 1991 | 37 | 1656 |
| 14 | 13 | **14** | 12 | **13** | **14** | Ambio | 1993 | 22 | 225 |
| 15 | | **11** | 7 | **8** | 11 | For. Ecol. Manage. | 2001 | 150 | 79 |
| 16 | 11 | 11 | 8 | 8 | 11 | Can. J. Forest Res. | 1992 | 22 | 1235 |
| 17 | 11 | 11 | 6 | **7** | 11 | For. Ecol. Manage. | 1994 | 69 | 299 |
| 18 | 10 | **11** | 2 | 2 | 11 | Agroforestry Forum | 1998 | 9 | 47 |
| 19 | 10 | 10 | 9 | **10** | 10 | For. Ecol. Manage. | 1997 | 94 | 149 |
| 20 | 6 | 6 | 9 | 9 | 9 | Photogramm. Eng. Rem. S. | 1990 | 56 | 1383 |

**Table 3**. Citations accruing to top 20 publications based on the union of both sources (GS and WoS), and excluding self-citations.

| Rank | GS | WoS | Max | Sum † | Cites | Exclude self-citations | Publication | Date | Vol | Page |
|---|---|---|---|---|---|---|---|---|---|---|
| 1 | 177 | 142 | 177 | | 177 | | Book: Modelling ... | 1994 | | |
| 2 | 58 | 52 | 58 | | 58 | | Forest Science | 1995 | 41 | 7 |
| 3 | 53 | 53 | 53 | | 53 | | Ecological Modelling | 1997 | 98 | 1 |
| 4 | 41 | 42 | 42 | | 42 | | For. Ecol. Manage. | 1995 | 71 | 267 |
| 5 | 42 | | 42 | | 42 | | Report: A Sustainable ... | 1999 | | |
| 6 | 27 | 40 | 40 | | 40 | | For. Ecol. Manage. | 1991 | 42 | 143 |
| 7 | 30 | 36 | 36 | | 36 | | For. Ecol. Manage. | 1989 | 27 | 245 |
| 8 | 30 | 32 | 32 | | 32 | | For. Ecol. Manage. | 1995 | 71 | 251 |
| 9 | 26 | 10 | 26 | | 26 | | For. Ecol. Manage. | 2003 | 172 | 229 |
| *10* | *14* | *13* | *14* | *27* | *22* | *20* | *Ambio* | *1993* | *22* | *225* |
| 11 | 15 | 19 | 19 | | 19 | 17 | J. Trop. For. Sci. | 1991 | 4 | 59 |
| 12 | 19 | 13 | 19 | | 19 | 10 | For. Ecol. Manage. | 2001 | 150 | 27 |
| 13 | 9 | 17 | 17 | | 17 | 16 | For. Ecol. Manage. | 1992 | 54 | 257 |
| 14 | 15 | 16 | 16 | | 16 | 15 | Forest Science | 1991 | 37 | 1656 |
| 15 | 11 | 8 | 11 | 19 | 14 | 10 | For. Ecol. Manage. | 2001 | 150 | 79 |
| 16 | 11 | 7 | 11 | 18 | 13 | | For. Ecol. Manage. | 1994 | 69 | 299 |
| 17 | 10 | 10 | 10 | 20 | 12 | | For. Ecol. Manage. | 1997 | 94 | 149 |
| 18 | 11 | 8 | 11 | 19 | 11 | | Can. J. Forest Res. | 1992 | 22 | 1235 |
| 19 | 11 | 2 | 11 | | 11 | | Agroforestry Forum | 1998 | 9 | 47 |
| 20 | 6 | 9 | 9 | | 9 | | Photogramm. Eng. Rem. S. | 1990 | 56 | 1383 |

† only for rows with Max(Scholar, ISI)<*h*-index and Sum(Scholar+ISI)>*h*-index.

In Table 3, only five papers fall into this category, and despite a relatively large change in the citations accruing to the *Ambio* paper, the *h*-index does not change. Finally, Table 3 also illustrates that it is a relatively simple matter to adjust for self-citations (because only 'threshold' publications need to be examined), and that despite the large number of co-author citations to one multi-author paper, the *h*-index changes only slightly, to 13. Clearly, the *h*-index is a robust indicator of published output in this instance. It is more difficult to verify its robustness for other researchers (i.e., hard to establish an error-free standard for verification without an intimate knowledge of the candidate publications), but unpublished trials based on the publications of colleagues suggest that the pattern illustrated here is representative.

The robustness of the *h*-index applies not only to individuals, but also to journals. *Forest Ecology and Management* (1995) is prominent in Table 3, so has been used to illustrate this in Table 4. This journal is ranked 1st by volume (20% of all forestry papers) and 5th by impact factor (out of 34 forestry journals), with an impact factor of 1.5 and a half-life of 5.9 (ISI 2004 JCR Science

Edition). A search (on 15 May 2006) for citations to articles in this journal appearing in 1995 generated 193 records with GS and 236 with WoS. When sorted, the naïve *h*-indices were 25 and 29 respectively. Correcting the more obvious errors reduced these to 185 and 195 records respectively, and did not change the *h*-indices, even though several records changed rank in both data sets. Table 4 illustrates the top 40 records, after correction, cross-matching and sorting. Combining both the GS and WoS records (by using the larger of the two for each publication; the union of the databases was not examined) indicates a *h*-index of 29. The *h*-index remained surprisingly stable, across two diverse sources, and despite a relatively large number of discrepancies in the raw data. It is interesting to observe that the discrepancy in the *h*-indices estimated from the two databases is about 15%, similar in magnitude to that observed by Cronin and Meho (2006) when comparing *h*-indices for faculty members derived from two different databases.

**Conclusion**

The integer nature of the *h*-index, its robustness to perturbations in the tails of the publication-citations distribution, and the ease of verifying, offer compelling reasons to favour a Hirsch-type index over an index based on total citations and total publications.

**Table 4**. Top 40 citations accruing to 1995 publications in *Forest Ecology and Management*. Bold entries denote the 'threshold' entry and *h*-index.

| Rank | GS | WoS | Max | 1st Author | Vol | Page |
|---|---|---|---|---|---|---|
| 1 | 139 | 206 | 206 | Dise N B | 71 | 153 |
| 2 | 90 | 89 | 90 | Aide T M | 77 | 77 |
| 3 | 78 | 70 | 78 | Brown I F | 75 | 175 |
| 4 | 75 | 63 | 75 | Verissimo A | 72 | 39 |
| 5 | 39 | 73 | 73 | Wright R F | 71 | 1 |
| 6 | 40 | 70 | 70 | Boxman A W | 71 | 7 |
| 7 | 24 | 54 | 54 | Emmett B A | 71 | 45 |
| 8 | 30 | 53 | 53 | Tietema A | 71 | 143 |
| 9 | 40 | 51 | 51 | Zimmerman J K | 77 | 65 |
| 10 | 36 | 48 | 48 | Larsen J B | 73 | 85 |
| 11 | 38 | 45 | 45 | Schowalter T D | 78 | 115 |
| 12 | 23 | 45 | 45 | Moldan F | 71 | 89 |
| 13 | 34 | 44 | 44 | Brandrud T E | 71 | 111 |
| 14 | 43 | 42 | 43 | Sheil D | 77 | 11 |
| 15 | 41 | 39 | 41 | Liu J G | 73 | 157 |
| 16 | 35 | 41 | 41 | Silva J N M | 71 | 267 |
| 17 | 20 | 41 | 41 | Gundersen P | 71 | 75 |
| 18 | 26 | 40 | 40 | Zou X M | 78 | 147 |
| 19 | 39 | 38 | 39 | Wright R F | 71 | 163 |
| 20 | 34 | 36 | 36 | Houllier F | 74 | 91 |
| 21 | 29 | 35 | 35 | Butterfield J | 79 | 63 |
| 22 | 25 | 35 | 35 | Brais S | 76 | 181 |
| 23 | 29 | 34 | 34 | Lurz P W W | 79 | 79 |
| 24 |  | 34 | 34 | Emmett B A | 71 | 61 |
| 25 | 30 | 32 | 32 | Soares P | 71 | 251 |
| 26 | 19 | 31 | 31 | Bredemeier M | 71 | 31 |
| 27 | 30 | 17 | 30 | Herrera J | 76 | 197 |
| 28 | 30 | 15 | 30 | Barros A C | 77 | 87 |
| 29 | 22 | **29** | **29** | Ranger J | 72 | 167 |
| 30 | 20 | 29 | 29 | Ashton M S | 72 | 1 |
| 31 | 14 | 28 | 28 | Degraaf R M | 79 | 227 |
| 32 | 19 | 27 | 27 | Madsen P | 72 | 251 |
| 33 | 23 | 26 | 26 | Butterfield R P | 75 | 111 |
| 34 | 21 | 26 | 26 | Wright R F | 71 | 133 |
| 35 | **25** | 24 | 25 | Maass J M | 74 | 171 |
| 36 | 25 | 23 | 25 | Iida S | 73 | 197 |
| 37 | 16 | 25 | 25 | Bosac C | 74 | 103 |
| 38 | 11 | 25 | 25 | Stuanes A O | 71 | 99 |
| 39 | 24 | 23 | 24 | Pausas J G | 78 | 39 |
| 40 | 15 | 24 | 24 | Bren L J | 75 | 1 |

## References

Agrawal, A.A. (2005) Corruption of journal impact factors. *TRENDS in Ecology and Evolution* 20, 4, 157.

Bakkalbasi, N., Bauer, K., Glover, J. & Wang, L. (2006) Three options for citation tracking: Google Scholar, Scopus and Web of Science. http://eprints.rclis.org/archive/00006080/ (26 June 2006).

Bensman, S.J. (in press) Garfield and Impact Factor: The Creation, Utilization, and Validation of a Citation Measure: Part 1, Creation and Utilization. ARIST 41, in press.

Bordons, M., Fernández, M.T. & Gómez, I. (2002). Advantages and limitations in the use of impact factor measures for the assessment of research performance. *Scientometrics* 53, 2, 195-206.

Bornmann, L. & Daniel, H.-D. (2005). Does the *h*-index for ranking of scientists really work? *Scientometrics*, 65, 3, 391–392.

Braun, T., Glänzel, W., & Schubert, A. (2005). A Hirsch-type index for journals. *The Scientist,* 19, 22, 8.

Butler, L. & Visser, M.S. (2006) Extending citation analysis to non-source items. *Scientometrics,* 66, 2, 327-343.

Cronin, B. & Meho, L. (2006). Using the *h*-index to rank influential information scientists. *Journal of the American Society for Information Science and Technology*, in press. 10.1002/asi.20354

Garfield, E. (2005) The agony and the ecstasy — the history and meaning of the journal impact factor: Presented at the International Conference on Peer Review and Biomedical Publication. http://garfield.library.upenn.edu/papers/jifchicago2005.pdf (26 June 2006)

Gehanno, J.-F., Darmoni, S.J. & Caillard, J.-F. (2005). Major inaccuracies in articles citing occupational or environmental medicine papers and their implications. *J Med Libr Assoc* 93(1):118-121.

Hirsch, J. (2005) An index to quantify an individual's scientific research output. *Proceedings of the National Academy of Sciences* (USA) 102, 46, 16569-16572.

Jennings, C. (1998). Citation data: the wrong impact? *Nature Neuroscience*, 1, 8, 641-642.

Karandikar, R.L., & Sunder, V.S. (2003) On the impact of impact factors. *Current Science*, 85, 3, 235.

Lange, L. (2002). The impact factor as a phantom: Is there a self-fulfilling prophecy effect of impact? *Journal of Documentation* 58, 2, 175-184.

Mannino, D.M. (2005). Impact Factor, Impact, and Smoke and Mirrors. *American Journal of Respiratory And Critical Care Medicine*, 171, 417-418.

Monastersky, R. (2005). The Number That's Devouring Science. *The Chronicle of Higher Education* 52, 8, A12, http://chronicle.com/free/v52/i08/08a01201.htm

Oppenheim, C. (2006) Using the h-index to rank influential British researchers in information science and librarianship. *Journal of the American Society for Information Science and Technology*, in press.

Redner, S. (1998). How Popular is Your Paper? An Empirical Study of the Citation Distribution. *European Physical Journal B*, 4, 2, 131 – 134.

Seglen, P.O. (1997). Why the impact factor of journals should not be used for evaluating research. BMJ 314, 497.